# The Gamma-ray Sky with *Fermi*

D. J. Thompson[a]

[a] *NASA Goddard Space Flight Center, Greenbelt, Maryland, 20771 USA*

**Abstract**

Gamma rays reveal extreme, nonthermal conditions in the Universe. The *Fermi Gamma-ray Space Telescope* has been exploring the gamma-ray sky for more than four years, enabling a search for powerful transients like gamma-ray bursts, solar flares, and flaring active galactic nuclei, as well as long-term studies including pulsars, binary systems, supernova remnants, and searches for predicted sources of gamma rays such as clusters of galaxies. Some results include a stringent limit on Lorentz invariance violation derived from a gamma-ray burst, unexpected gamma-ray variability from the Crab Nebula, a huge gamma-ray structure in the direction of the center of our Galaxy, and strong constraints on some Weakly Interacting Massive Particle (WIMP) models for dark matter.

## 1. Introduction

Cosmic gamma rays are signatures of high-energy particle interactions taking place in astrophysical settings. For this reason, gamma-ray telescopes enable studies of extreme, nonthermal phenomena in the Universe. Space is largely transparent to gamma rays in the GeV energy range and below, but the Earth's atmosphere is not. Direct detection of these gamma rays therefore requires space-based instruments.

## 2. *The Fermi Gamma-ray Space Telescope*

A successor to the successful *Compton Gamma Ray Observatory* that operated from 1991-2000, the *Fermi Gamma-ray Space Telescope* was launched by NASA into low-Earth orbit in June 2008 [1]. *Fermi* carries two scientific instruments, both built by international partnerships: (1) the Large Area Telescope (LAT) is a pair-production detector operating in the energy range from about 20 MeV to more than 300 GeV [2]; and (2) the Gamma-ray Burst Monitor (GBM) is a set of inorganic scintillators for the energy range 8 keV to 40 MeV [3]. Compared to most telescopes, the *Fermi* instruments have extremely large fields of view, 2.4 sr and 9 sr respectively. The observatory is usually operated in a scanning mode with the instruments pointed away from the Earth, allowing a survey of the full sky every two orbits, or about 3 hours.

Four years of operation have produced a broad range of scientific results from the *Fermi* instruments. This brief review is far from comprehensive. It highlights some observations particularly related to particle acceleration and interaction, while simply mentioning others. Following an overview of the gamma-ray sky, the discussion is ordered by source distance from the Earth.

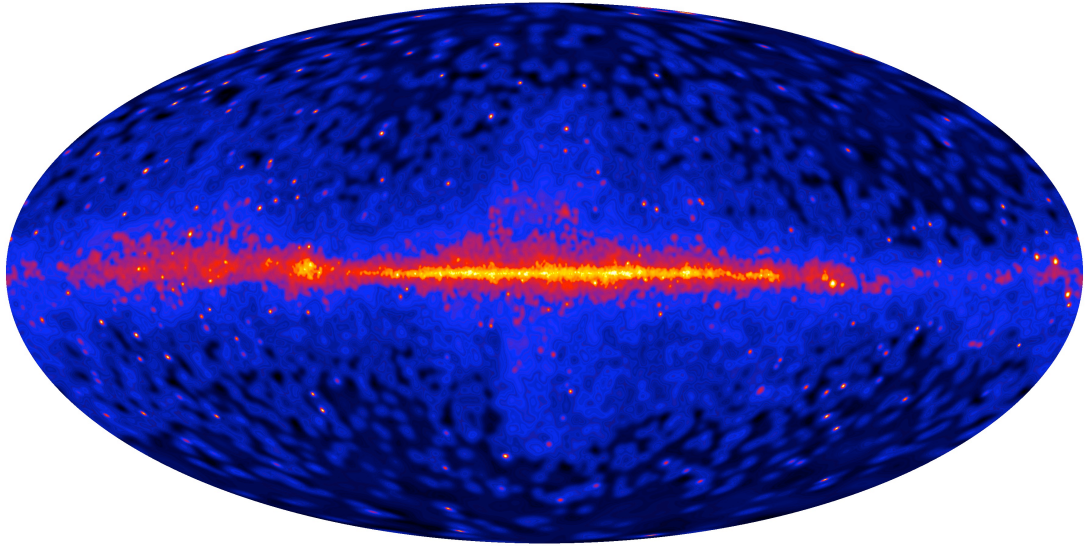

Fig. 1. Brightness map of the sky at energies above 10 GeV, shown in Galactic coordinates, adaptively smoothed to reduce statistical fluctuations in regions of low gamma-ray intensity. Figure courtesy of the *Fermi* LAT Collaboration.

### 3. The Gamma-ray Sky

Figure 1 shows a smoothed gamma-ray counts map of the entire sky at energies above 10 GeV, based on four years of data from the *Fermi* LAT. The projection uses Galactic coordinates, with the plane of the Milky Way as the horizontal axis and the origin representing the direction to the Galactic Center. The gamma-ray sky is not completely dark in any direction, because it has an isotropic background glow from unresolved extragalactic sources. The Galactic plane is a bright band, resulting primarily from cosmic-ray particle interactions with the interstellar gas and photon fields rather than from individual gamma-ray-producing objects. Many discrete gamma-ray sources are also visible as bright spots, seen in all directions.

Although the Galactic gamma-ray emission has been long known, a new and unexpected feature appears in this figure in the form of the large lobes extending above and below the Galactic Center direction. Assuming these "Fermi Bubbles" are at the distance of the Galactic Center, their energy spectrum and well-defined spatial shape suggest that they represent a huge reservoir of energetic particles possibly produced during some previous era of activity in the core region of the Milky Way [4].

### 4. Gamma-ray Sources

Seen against the diffuse emission from the Milky Way and the isotropic background are individual gamma-ray sources. The Second *Fermi* LAT Catalog (2FGL), based on the first two years of LAT data, contains 1873 sources [5], of which all but 12 are point sources within the spatial resolution of the LAT.

### 5. Gamma-ray Sources in our Galaxy

The GBM and the LAT have detected gamma rays from Solar System objects, including the Sun, the Moon, and even thunderstorms on Earth. The largest classes of relatively nearby gamma-ray sources, however, are pulsars (plus pulsar wind nebulae) and supernova remnants.

*5.1. Pulsars and Pulsar Wind Nebulae*

Rapidly rotating, magnetized neutron stars generate strong electric fields that accelerate particles to high enough energies that they can produce gamma rays, principally through curvature radiation. The

LAT has now seen pulsed gamma radiation from over 100 of these pulsars[*]. Key features of these gamma-ray pulsars are:

- About 1/3 are seen only in gamma rays and not in radio or other wavelength bands, implying that the gamma-ray beam is broad and sweeps out much of the sky [6] [7] [8]. These pulsars therefore represent a relatively unbiased sample of nearby neutron stars.
- Another 1/3 are millisecond pulsars, which are thought to be old, "recycled" neutron stars that were parts of binary systems. Angular momentum transfer during the binary phase spun up the neutron star to produce the very short rotation periods [9].
- Multiwavelength cooperation has been important to pulsar studies. Over 40 new millisecond pulsars have been discovered by radio astronomers searching locations of unidentified LAT sources, and many of these are now known to be gamma-ray pulsars as well [10] [11] [12]. The radio community is particularly interested in stable millisecond pulsars, because an array of such pulsars can in principle be used to detect gravitational radiation [13].
- These gamma-ray pulsars are highly efficient particle accelerators. For many, at least 10% of the neutron star's rotational energy loss is being converted into pulsed gamma radiation [14].

In addition to the direct particle acceleration in the vicinity of the neutron star, pulsars produce strong particle outflows into their environments, creating pulsar wind nebulae, whose shock fronts also accelerate particles. Interactions of these high-energy particles with the ambient medium can make such pulsar wind nebulae gamma-ray sources [15]. The best-known of these is the Crab Nebula, which surprised the astrophysical community by producing strong, fast outbursts of gamma radiation seen by both the Italian gamma-ray telescope *AGILE* [16] and the *Fermi* LAT [17], implying localized particle acceleration to PeV energies. The acceleration mechanism remains a puzzle, since shock acceleration seems too slow a process to produce the characteristics of the flares that have been seen.

---

[*] Information about the known gamma-ray pulsars can be found at
https://confluence.slac.stanford.edu/display/GLAMCOG/Public+List+of+LAT-Detected+Gamma-Ray+Pulsars.

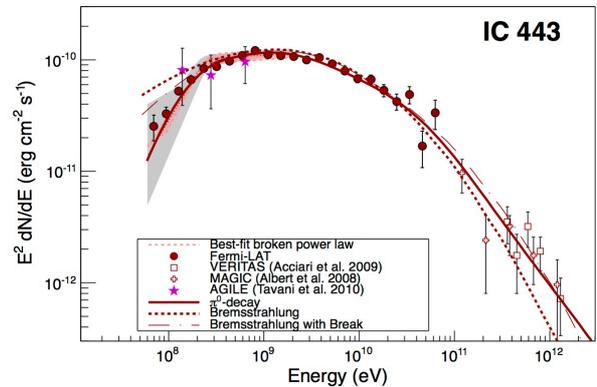

Fig. 2. Gamma-ray spectral energy distribution of the IC 443 supernova remnant [24].

*5.2. Supernova Remnants*

Supernovae and their remnants have been circumstantially identified as the sources of the cosmic-ray particles that fill space in our Galaxy [18]. Gamma-ray studies of supernova remnants (SNR) are now providing strong evidence for this conclusion, as the particles accelerated in these objects interact with the local surrounding medium to produce gamma rays. Two types of SNR have been seen as gamma-ray sources: (1) Young SNR like Tycho [19] and Cassiopeia A [20], with ages of a few kyr; and (2) Older SNR, with ages tens of kyr, that are close to large molecular clouds, such as W28 [21] and W51C [22]. These molecular clouds provide target material for cosmic-ray interactions, and the cosmic rays have had more time to escape from these SNR, making them typically more luminous in GeV gamma rays than the younger ones [23].

The primary challenge has been in establishing that these SNR accelerate protons, which represent the bulk of the Galactic cosmic rays. For a number of these SNR, modeling has favored models in which the gamma rays are produced through hadronic interactions, principally the production of $\pi^0$ mesons, which quickly decay to gamma rays in the LAT energy range. The most convincing cases are the two brightest gamma-ray SNR, W44 and IC443 [24]. Fig. 2 shows the spectral energy distribution of IC443, and the spectrum of W44 is similar. The rollover at energies below 100 MeV is very hard to explain by any mechanism other than $\pi^0$ decay. Although a sample of two does not prove that all or even most cosmic rays are produced in SNR, it does establish that at least some SNR are accelerating protons to cosmic-ray energies.

*5.3. Other Galactic Gamma-ray Sources*

Four high-mass binary systems are found among the *Fermi* LAT sources, with orbital periods ranging from 4.8 hours (Cygnus X-3 [25]) to 3.4 years (the PSR B1259-58 system [26] [27]). Although each of these binaries contains a massive normal star, the nature of the companion star is unknown in all of them except PSR B1259-58, where the companion is a neutron star. The nature of the particle acceleration and interaction processes in these systems remains the subject of ongoing studies.

## 6. Extragalactic Gamma-ray Sources

While the vast majority of the gamma-ray sources in the Milky Way are relatively persistent (steady or having regular periodic variability), the extragalactic gamma-ray sky is characterized by extreme variability.

*6.1. Active Galactic Nuclei*

Over half the sources in the *Fermi* LAT catalog are associated with active galactic nuclei (AGN), generally thought to be powered by supermassive black holes. Most of these are identified with the AGN subclass known as blazars, including flat-spectrum radio quasars and BL Lacertae objects, in which a powerful jet of particles and photons is directed nearly toward the Earth. The Doppler boosting in the jet helps explain the extremely high apparent luminosities and rapid time variations seen in some gamma-ray blazars.

An example of the extreme variability seen by the LAT is shown in Fig. 3. The daily average flux from 3C454.3 varies by over two orders of magnitude,

---
[†] Automatically generated daily and weekly flux histories for a number of LAT sources, including 3C 454.3, are available at http://fermi.gsfc.nasa.gov/ssc/data/access/lat/msl_lc/. Please note that these data are preliminary and not intended for detailed scientific analysis.

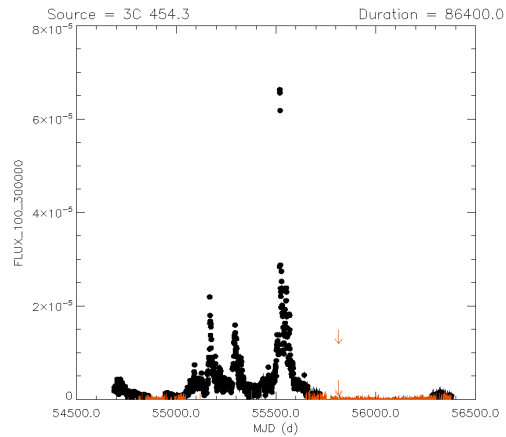

Fig. 3. Average daily flux (E>100 MeV) seen by the *Fermi* LAT for blazar 3C 454.3 from the start of the *Fermi* science mission (2008 August) until 2013 March. Red arrows (primarily clustered along the x-axis with the arrowheads invisible) indicate days in which the blazar was not detected[†].

with variability seen on time scales as short as three hours [28] [29].

All the AGN seen by *Fermi* have emission that extends across much of the electromagnetic spectrum, often from the radio to TeV gamma-ray bands, e.g. [30]. The lower-energy radiation (radio to optical) appears to be synchrotron radiation from a population of energetic electrons, while the higher-energy emission is likely to be produced by inverse Compton scattering from some photon field by the same population of electrons, although hadronic jet models cannot be ruled out.

The nearly continuous sky survey in gamma rays by *Fermi*, combined with the wealth of available multiwavelength data, offer the possibility of detailed spectral and temporal studies of AGN. Results have revealed a complex situation, with modeling of the broadband spectral energy distribution often favoring production of the gamma rays not far from the central black hole (e.g. [31]), while time variability analysis often indicates that the gamma-ray origin lies farther from the central engine (e.g. [32]). A recent review of this evolving field is given by [33].

Blazars seen by the *Fermi* LAT have also been used to probe the Extragalactic Background Light, which comes from all the stars that have ever existed. Gamma rays from distant AGN interact with optical and ultraviolet photons through $\gamma + \gamma \rightarrow e^+ + e^-$, producing an absorption feature that has been seen in their LAT spectra [34]. The result places strong constraints on the existence of an unseen population of stars in the early Universe.

## 6.2. Gamma-ray Bursts

Gamma-ray bursts (GRBs) have been called the most powerful explosions since the Big Bang. These intense flashes of gamma rays, lasting from a fraction of a second to minutes (plus even longer afterglows), are thought to result from extreme types of supernovae or from the mergers of compact object binaries like neutron stars and black holes. With its large field of view and collecting area, the *Fermi* GBM currently produces the largest number of GRB detections, while the *Fermi* LAT detects the smaller number of those bursts that have higher-energy emission. The combination of GBM and LAT simultaneous burst observations shows that many GRBs have energy spectra that become steeper or have cutoffs at higher energies [35].

The fact that GRBs are short, have a broad range of photon energies, and come from cosmological distances implies that they are useful for direct tests of Lorentz invariance violation, predicted by some models of quantum gravity [36]. GRB 090510 originated from a galaxy with redshift 0.9, implying that the gamma rays traveled for about 7 billion years. The burst included a 31 GeV photon, which arrived less than 1 s after the onset of the burst at lower energies [37]. Under conservative assumptions, the lack of more dispersion in photon arrival times sets a constraint on the quantum gravity mass scale that exceeds the Planck mass, $1.22 \times 10^{19}$ GeV.

## 7. Implications of Unseen Gamma-Ray Sources

What is not seen by a telescope can also be important. Two examples from *Fermi* illustrate this point.

### 7.1. Clusters of Galaxies

Clusters of galaxies are the largest gravitationally bound structures in the Universe, and many of them are known to produce strong shocks and winds that can accelerate particles, as seen by nonthermal radio and X-ray emission. Detectable levels of gamma radiation were predicted from a number of clusters, e.g. [38]. Clusters also contain concentrations of dark matter, whose particle annihilations could potentially produce observable gamma-ray signals, e.g. [39]. No clusters of galaxies appear in the *Fermi* LAT catalogs, however, and analyses targeted to such clusters have produced only upper limits [40] [41]. Re-evaluation of the predictions seems necessary.

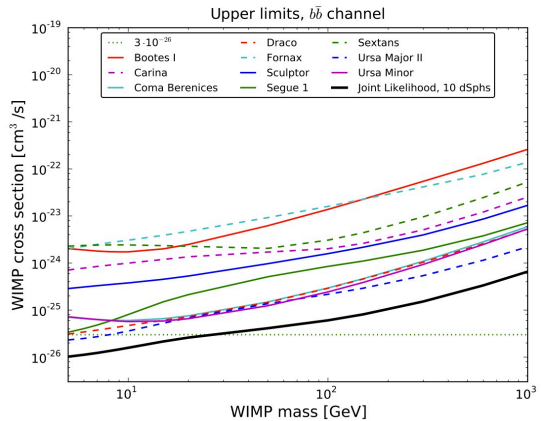

Fig. 4. Upper limits (95% confidence) on a velocity-averaged WIMP annihilation cross section for selected dwarf spheroidal galaxies and for the joint likelihood analysis for annihilation into the bb-bar final state [44]. A generic cross section ($\sim 3 \times 10^{-26}$ cm$^3$s$^{-1}$) is plotted (horizontal dotted line) as a reference.

### 7.2. Dwarf Spheroidal Galaxies

Among the many ways cosmic gamma rays can be used to search for indirect evidence of dark matter, detection of dwarf spheroidal galaxies would be one of the most robust methods. These small satellite galaxies of the Milky Way are dominated by dark matter, have little gas, and show low rates of star formation [42] [43], making them unlikely to be gamma-ray sources by the conventional processes seen in the diffuse emission from the Galaxy or in Galactic sources.

No dwarf spheroidals have been seen in the *Fermi* LAT data. Conversion of the upper limits on gamma-ray fluxes to upper limits on dark matter Weakly Interacting Massive Particle (WIMP) properties clearly involves model assumptions about the particle annihilation channels. With reasonable assumptions, various authors have produced upper limits on these particle properties, e.g. [44] [45]. As seen in Fig. 4, the combined upper limits from a number of dwarf spheroidals give a value, for assumed WIMP masses below a few tens of GeV, below the $\sim 3 \times 10^{-26}$ cm$^3$s$^{-1}$ average product of cross section and particle speed that would be expected if such particles were to explain the known dark matter in the Universe.

## 8. Conclusion

Less than halfway through its nominal 10-year mission, the *Fermi Gamma-Ray Space Telescope* continues to produce a broad range of results on many types of energetic phenomena. Particularly at the highest energies for the LAT, improvements in analysis methods and increases in photon counts are enhancing the source measurement capabilities at a rate almost linear with time. The GBM has changed operating modes to deliver time-tagged information about each photon, making it more sensitive, particularly to short GRBs that will be of particular interest in the era of Advanced LIGO/VIRGO searches for gravitational wave signals.

Because *Fermi* is a facility-class mission, all the gamma-ray data from the instruments are made public immediately, along with analysis software and detailed documentation, through the *Fermi* Science Support Center[‡], which also offers help for anyone interested in using *Fermi* data. In addition, the instrument teams welcome cooperative scientific efforts. More discoveries are waiting.

## Acknowledgments

Results from the *Fermi Gamma-Ray Space Telescope* are the product of many years' efforts by the LAT and GBM instrument teams, the project staff, and the scientific user community. I extend my thanks to all of them for making this summary possible.

## References

[1] J. E. McEnery, P. F. Michelson, W. S. Paceisas, & S. Ritz, Optical Engineering, 51 (2012) 011012.
[2] W. B. Atwood et al., ApJ. 697 (2009) 1071.
[3] C. Meegan et al., ApJ. 702 (2009) 791.
[4] M. Su, T. R. Slatyer, & D. P. Finkbeiner, ApJ 724 (2010) 1044.
[5] P. L. Nolan et al, ApJ Supp. 199 (2012) 31.
[6] A. A. Abdo et al., Science 325 (2009) 840.
[7] P. M. Saz Parkinson et al., ApJ 725 (2010) 571.
[8] H. J. Pletsch et al., ApJ 744 (2012) 105.
[9] A. A. Abdo et al., Science 325 (2009) 848.
[10] S. M. Ransom et al., ApJ 727 (2011) L16.
[11] I. Cognard et al., ApJ 732 (2011) 47.
[12] M. J. Keith et al., MNRAS 414, (2011) 1292.
[13] R. N. Manchester, in American Institute of Physics Conference Series, vol. 1357, 2011, 65.
[14] A. A. Abdo, A. A. et al., ApJ Supp 187 (2010) 460.
[15] M. Ackermann et al., ApJ 726 (2011) 35.
[16] M. Tavani et al., Science 331 (2011) 736.
[17] R. Buehler et al., ApJ,749, (2012) 26.
[18] V. I. Ginzburg & S. I. Syrovatskii, The Origin of Cosmic Rays, Macmillan, New York, 1964.
[19] F. Giordano et al., ApJ 744 (2012) L2.
[20] A. A. Abdo et al., ApJ 710 (2010) L92.
[21] A. A. Abdo et al., ApJ 718 (2010) 348.
[22] A. A. Abdo et al., ApJ 706 (2009) L1.
[23] D. J. Thompson, L. Baldini, & Y. Uchiyama, Astropart. Phys. 39 (2012) 22.
[24] M. Ackermann et al., Science 339 (2013) 807.
[25] A.A. Abdo et al., Science 326 (2009) 1512.
[26] P. H. T. Tam et al., ApJ 736 (2011) L1.
[27] A.A. Abdo et al., ApJ 736 (2011) L11.
[28] M. Ackermann et al., ApJ 721 (2010) 721.
[29] A. A. Abdo et al., ApJ 733 (2011) L26.
[30] A.A. Abdo et al., 2011, ApJ 736 (2011) 131.
[31] G. Ghisellini & F. Tavecchio, MNRAS 397 (2009) 985.
[32] A. P. Marscher et al., ApJ 710 (2010) 126.
[33] J. Finke, Proc. 2012 Fermi Symposium, eConf C121028 (2013) arXiv:1303.5095v1.
[34] M. Ackermann et al., Science 338 (2012) 1190.
[35] M. Ackermann et al., ApJ 754 (2012) 121.
[36] G. Amelino-Carmelia et al., Nature 393 (1998) 763.
[37] A. A. Abdo et al., Nature 462 (2009) 331.
[38] P. Blasi, S. Gabici, & G. Brunetti, International Journal of Modern Physics A 22, (2007) 681.
[39] T. E. Jeltema, J. Kehayias, & S. Profumo, Phys. Rev. D 80 (2009) 023005.
[40] M. Ackermann et al., ApJ 717 (2010) L71.
[41] M. Ackermann et al., JCAP 05 (2010) 025.
[42] M. Mateo, Ann. Rev. Astron. Astrophys. 36 (1998) 435.
[43] J. Grcevich & M. E. Putman, Astrophys. J. 696 (2009) 385.
[44] M. Ackermann et al., Phys. Rev. Lett. 107 (2011) 241302.
[45] A. Geringer-Sameth & S. M. Koushiappas, Phys. Rev. Lett. 107 (2011) 241303.

---

[‡] Web access for the data and all supporting information is available at http://fermi.gsfc.nasa.gov/ssc/.